\documentclass[a4paper,11pt]{article}
\usepackage{pos}
\usepackage{subfigure}
\usepackage{graphicx}
\usepackage{amsmath} 
\usepackage{relsize}%
\usepackage[utf8]{inputenc} 
\usepackage{bbm}
\usepackage{xfrac}
\usepackage{fancybox}
\usepackage{xkeyval}
\usepackage{ulem}

\newcommand{\bea}{\begin{eqnarray}}
\newcommand{\eea}{\end{eqnarray}} 
\usepackage{environ}
\NewEnviron{myequation}{%
\begin{equation}
\scalebox{1.5}{$\BODY$}
\end{equation}
}
\definecolor{orange}{rgb}{1,0.5,0}
\newcommand{\beq}{\begin{equation}}
\newcommand{\eeq}{\end{equation}}

\renewcommand{\l}{\lambda}

\newcommand{\p}{\phi}
\renewcommand{\a}{\alpha}

\newcommand{\tr}{\text{Tr}}
\newcommand{\hk}{\hat{k}}

\newcommand{\bx}{\mathbf{x}}
\newcommand{\by}{\mathbf{y}}

\renewcommand{\p}{\psi}
\newcommand{\pb}{\overline{\psi}}
\newcommand{\vx}{\bx}
\newcommand{\vy}{\by}

\newcommand{\n}{\nu}
\newcommand{\m}{\mu}

\newcommand{\dg}{\dagger}
\newcommand{\non}{\nonumber}

\newcommand{\ra}{\rightarrow}

\usepackage{ulem}
\usepackage{xcolor}
\definecolor{olive}{rgb}{0.3, 0.4, .1}
\definecolor{fore}{RGB}{249,242,215}
\definecolor{back}{RGB}{51,51,51}
\definecolor{title}{RGB}{255,0,90}
\definecolor{dgreen}{rgb}{0.,0.5,0.}
\definecolor{gold}{rgb}{1.,0.84,0.}
\definecolor{JungleGreen}{cmyk}{0.99,0,0.52,0}
\definecolor{BlueGreen}{cmyk}{0.85,0,0.33,0}
\definecolor{RawSienna}{cmyk}{0,0.72,1,0.45}
\definecolor{Magenta}{cmyk}{0,1,0,0}

\begin{document}
\title{Excitations of isolated static charges in q=2 Abelian Higgs theory}

\author*{Kazue Matsuyama}

\affiliation{Physics and Astronomy Department, San Francisco State University,\\
  1600 Holloway Ave., San Francisco CA USA}

\emailAdd{kazuem@sfsu.edu}

\abstract{In this report, I discuss lattice Monte Carlo evidence of excitations of isolated static charges in q=2 Abelian Higgs theory. The localized excitations are excited states of the interacting fields surrounding the static charges.}

\FullConference{%
 The 38th International Symposium on Lattice Field Theory, LATTICE2021
  26th-30th July, 2021
  Zoom/Gather@Massachusetts Institute of Technology
}


\maketitle

\section{Introduction}
In quantum chromodynamics, a spectrum of localized quantum excitations of the surrounding field exists for a static quark and antiquark pair in the confining phase of a pure gauge theory. This is the spectrum of localized quantum excitations of the color electric field associated with the pair of color charges which forms a flux tube between the quark and antiquark pair. This flux tube indeed exists in a number of vibrational modes.

On the other hand, in ordinary quantum electrodynamics, any disturbance of the field surrounding a static charge can be viewed as 
the creation of some set of photons superimposed on a Coulombic background. Therefore, there are no stable localized excitations of the electromagnetic field in QED. However, one may ask if there could be such a spectrum of localized excitations in the Higgs phase in gauge Higgs theories, which are interacting but non-confining. 

These proceedings are the based on the work reported in \cite{Matsuyama2021}.

\section{Theoretical model and Monte Carlo simulations}
In order to explore our idea, we consider an interacting gauge Higgs theory, namely the $q=2$ Abelian Higgs model,
\beq
 S  = - \beta \sum_{plaq}  \mbox{Re}[U_\m(x)U_\n(x+\hat{\m})U_\m^*(x+\hat{\n}) U^*_\n(x)]  \\
             - \gamma \sum_{x,\m}  \mbox{Re}[\phi^*(x)U^2_\m(x) \phi(x+\widehat{\m})] \ ,
\eeq
where the scalar field has charge $q=2$ as do Cooper pairs. For simplicity of our calculations, we impose a unimodular constraint  ${\phi^*(x) \phi(x) = 1}$. Our motivation is that this model is a relativistic generalization of the Landau-Ginzburg effective model of superconductivity. 

For constructing our model, we consider physical states containing a static fermion and anti-fermion sitting at sites x and y. And we look for stable
localized excitations of the U(1) gauge field and Higgs field surrounding a static charge. In fact, such excitations have been reported in SU(3) gauge Higgs theory \cite{Jeff2020} and in chiral U(1) gauge Higgs theory by Jeff Greensite \cite{Jeff2021}.

With this theoretical model, we compute the energy excitations above the ground state of the Higgs + U(1) gauge field surrounding a pair of static charges via lattice Monte Carlo simulations, and in our calculation the excited state is stable if the energy of the excited state above the ground state is less than the photon mass.

To identify such localized excitations in our Monte Carlo simulations, let us first consider a physical state for a static fermion and anti-fermion pair at 
sites x and y, each of $\pm 2$ electric charge: 
\beq
        |\Phi_\a(R)\rangle = Q_\a(R) |\Psi_0\rangle \ , 
 \eeq
where $\Psi_0$  is the vacuum state, and 
 \beq 
 Q_\a(R) = [\pb(\vx) \zeta_\a(\vx)] ~\times~  [\zeta^*_\a(\vy) \p(\vy)] \ .
 \eeq
Here the $\pb, \p$ are operators creating double-charged static fermions of opposite charge 
transforming as ${\p(x) \ra e^{2i\theta(x)} \p(x)}$, and the $\{\zeta_\a(x)\}$  are a set of operators,
 which may depend on some (possibly non-local) combination of the Higgs and gauge fields,
 also transforming as ${\zeta(x) \ra e^{2i\theta(x)} \zeta(x)}$ under a gauge transformation, where the link variables transform as 
 \beq
    U_\m(x) \ra  \exp(i\theta(x)) U_\m(x) \exp(-\theta(x+\hat{\mu})). 
  \eeq 
Then possible choices for $\zeta$ operators are first of all the Higgs field $\phi(x)$ and another set is given by eigenstates $\zeta=\xi_\a$ of the covariant Laplacian, where 
 \beq
   (-D_iD_i)_{xy}\xi_\a({\vy};U) = \lambda_\a\xi_\a({\vx};U)
 \eeq 
and 
\beq
   (-D_iD_i)_{xy} = \sum_{k=1}^3 [2\delta_{\vx\vy} - U^2_k({\vx}) \delta_{{\vy, \vx} + \hk} - U^{* 2}_k ({\vx} - \hk)\delta_{{\vy, \vx} - \hk}] \ .
 \eeq 
Because the covariant Laplacian depends only on the squared link variable, the $\xi_\a(x;U)$, which we have elsewhere referred to as  ``pseudomatter'' fields, transform like $q=2$ charged matter fields, with the one difference that, unlike matter fields, they do not transform under a global (constant) gauge transformation.  Pseudomatter fields depend nonlocally on the gauge fields, and the low-lying eigenstates and eigenvalues of the covariant Laplacian, which is a sparse matrix, can be computed numerically via the Arnoldi algorithm.

In our calculations then, we make use of the four lowest-lying Laplacian eigenstates and the Higgs field to construct the $\Phi_\a$,
\beq
  \zeta_i(x) =  \left\{ \begin{array}{cl} 
                     \xi_i(x) &  i=1,2,3,4 \cr
                     \phi(x)  &  i=5 \end{array} \right.
\eeq
In general the five states $\Phi_\a(R)$ are non-orthogonal at finite $R$.  Of course $\phi(x)$ is a $q=2$ matter field, rather than pseudomatter field. We express the operator $Q_\a$ in terms of a non-local operator $V_\a(\vx,\vy;U)$
 \bea
  Q_\a(R) &=& \pb(\vx) V_\a(\vx,\vy;U) \p(\vy) \\
 V_\a(\vx,\vy;U) &=&   \zeta_\a(\vx;U) \zeta^*_\a(\vy;U)   \ , 
 \eea 
and define the Euclidean time evolution operator of the lattice abelian Higgs model, $\mathcal{T} = e^{-(H-\mathcal{E}_0)}$, which is the transfer matrix multiplied by a constant $e^{\mathcal{E}_0}$ where $\mathcal{E}_0$ is the vacuum energy, evolving states for one unit of discretized time.  

To compute the rescaled transfer matrix, we define $\mathcal{[T]}_{\alpha\beta}$ as the matrix element in the five non-orthogonal states $\Phi_\alpha$, with the matrix of overlaps, $\mathcal{[O]}_{\alpha\beta}$, of such states.
\bea
  [\mathcal{T}]_{\alpha\beta} &=& \langle \Phi_\alpha | e^{-(H-\mathcal{E}_0)} | \Phi_\beta \rangle  = \langle Q_\alpha^\dagger(R,1) Q_\beta(R,0) \rangle  \\
     \left[ O \right]_{\alpha\beta} &=&  \langle \Phi_\alpha | \Phi_\beta \rangle  = \langle Q_\alpha^\dg(R,0) Q_\beta(R,0) \rangle 
\eea 
 We obtain the five orthogonal eigenstates of $\mathcal{[T]}_{\alpha\beta}$ in the subspace of Hilbert space spanned by the $\Phi_\alpha$ by solving the generalized eigenvalue problem.
  \beq
      [T]_{\alpha\beta} \upsilon_\beta^{(n)} = \lambda_n [O]_{\alpha\beta}\upsilon_\beta^{(n)} \ ,
 \eeq
with eigenstates,
  \beq
   \Psi_n(R) = \sum_{\alpha=1}^5 \upsilon^{(n)}_\alpha \Phi_\alpha(R) \ 
  \eeq
and ordered such that $\lambda_n$ decreases with $n$.

Consider evolving the states $\Psi_n$ in Euclidean time,
\bea
         \mathcal{T}_{nn}(R,T) &=& \langle \Psi_n | e^{-(H-\mathcal{E}_0)T} | \Psi_n \rangle  \\
                                 &=& \upsilon^{*(n)}_\alpha \langle \Phi_\a | e^{-(H-\mathcal{E}_0)T}  | \Phi_\beta \rangle \upsilon^{(n)}_\beta   \\
                                 &=& \upsilon^{*(n)}_\alpha  \langle Q_\alpha^\dagger(R,T) Q_\beta(R,0) \rangle \upsilon^{(n)}_\beta  \ , 
 \eea
where Latin indices indicate matrix elements with respect to the $\Psi_n$ rather than the $\Phi_\alpha$, and there is a sum
over repeated Greek indices.

To calculate this expression, we define timelike $q=2$ Wilson lines of length T,
\beq
      P(\vx,t,T) = U^2_0(\vx,t) U^2_0(\vx,t+1)...U^2_0(\vx, t +T-1) \ .
\eeq
After integrating out the massive fermions, whose worldlines lie along timelike Wilson lines, we have
\beq
       \langle Q_\alpha^\dagger(R,T) Q_\beta(R,0) \rangle \\
       = \langle \tr[V^\dagger_\alpha(\vx,\vy;U(t+T)) P^\dagger(\vx,t,T) V_\beta(\vx,\vy;U(t)) P(\vy,t,T)]  \rangle \ .
\eeq
On general grounds, $\mathcal{T}_{nn}(R,T)$ is a sum of exponentials
\bea
      \mathcal{T}_{nn}(R,T)  &=&   \langle \Psi_n(R) | e^{-(H-\mathcal{E}_0)T} | \Psi_n(R) \rangle  \\
                            &=&   \sum_j |c^{(n)}_j(R)|^2 e^{-E_j(R) T} \ , \non
\eea
where $c_j^{(n)}(R)$ is the overlap of state $\Psi_n(R)$ with the j-th energy eigenstate of the abelian Higgs theory containing a static fermion-antifermion pair at separation $R$, and $E_j(R)$ is the corresponding energy eigenvalue minus the vacuum energy.

\section{Numerical Results}
\begin{figure}[htb]
\center 
\includegraphics[scale=0.5]{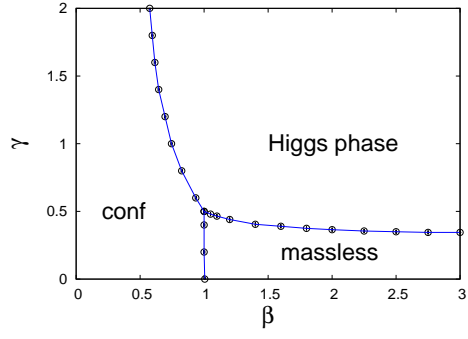} 
\caption{Thermodynamic phase diagram for $q=2$ Abelian Higgs model}
\label{fig1}
\end{figure}

We are interested in determining $E_n(R)$ in the Higgs phase and, because the calculation involves fitting exponential decay, we would like both the mass of the photon and the energies $E_n(R)$ to be not much larger than unity in lattice units.  For this reason we choose to work at the edge of the phase diagram shown, just above the massless-to-Higgs transition line at $\beta=3, \gamma=0.5$ in Fig.\ \ref{fig1}.

We compute the photon mass from the gauge invariant on-axis plaquette-plaquette correlator with the same
$\mu$ $\nu$ orientation
\bea
         G(R) = \bigg\langle {\rm Im}[U_\m(x)U_\n(x+\hat{\m})U_\m^*(x+\hat{\n}) U^*_\n(x)]
       \times {\rm Im}[U_\m(y)U_\n(y+\hat{\m})U_\m^*(y+\hat{\n}) U^*_\n(y)] \bigg\rangle \ ,
\eea
where $y = x + R \hat{k}$, and $\hat{k}$ is a unit vector orthogonal to the $\hat{\mu},\hat{\nu}$ directions.

\begin{figure}[htb]
\center
\includegraphics[scale=0.5]{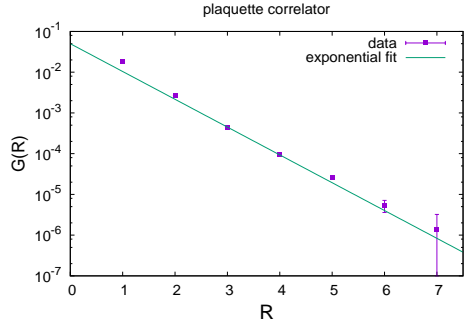}   
\caption{On-axis plaquette-plaquette correlator as a function of distance R for $\beta=3, \gamma=0.5$}
\label{fig2}
\end{figure}

The result for the $\beta=3, \gamma=0.5$ parameters is shown in Fig.\ \ref{fig2}.
From an exponential fit, disregarding the initial points, we find a photon mass of $m_\gamma = 1.57(1)$ in lattice units.  
Data was obtained on a $16^4$ lattice with 1,600,000 sweeps and data taken every 100 sweeps.  We have checked that if the
calculation is done just below the transition, in the massless phase, then $G(R)$ is fit quite well by a $1/R^4$ falloff, as expected. 

The energies $E_n(R)$ for $n=1,2$ are also obtained by fitting the data for $\mathcal{T}_{nn}(R,T)$ vs.\ $T$, \\ at each $R$, to an exponential falloff. 
\begin{figure}[htb]
\center
\subfigure[~]
{
\label{fig3}
\includegraphics[scale=0.5]{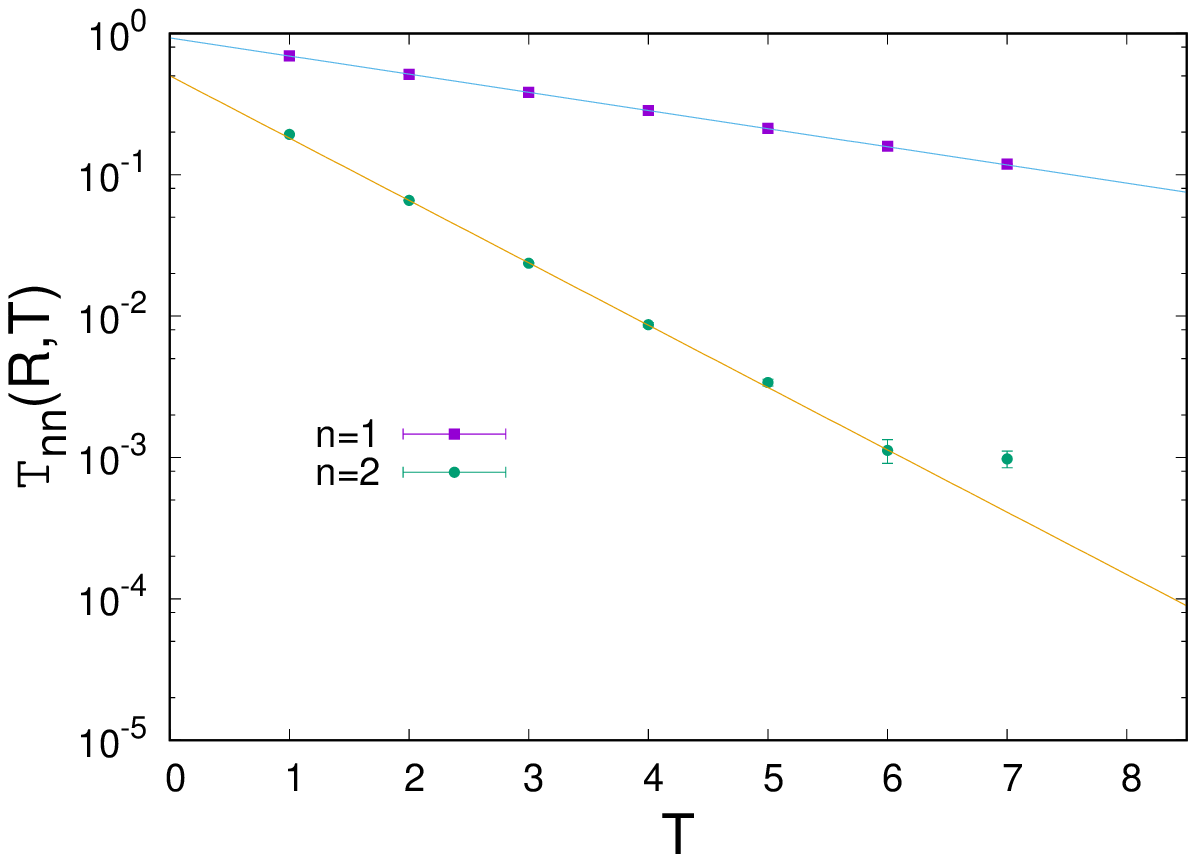}
}
\subfigure[~]
{
\label{fig4}
\includegraphics[scale=0.5]{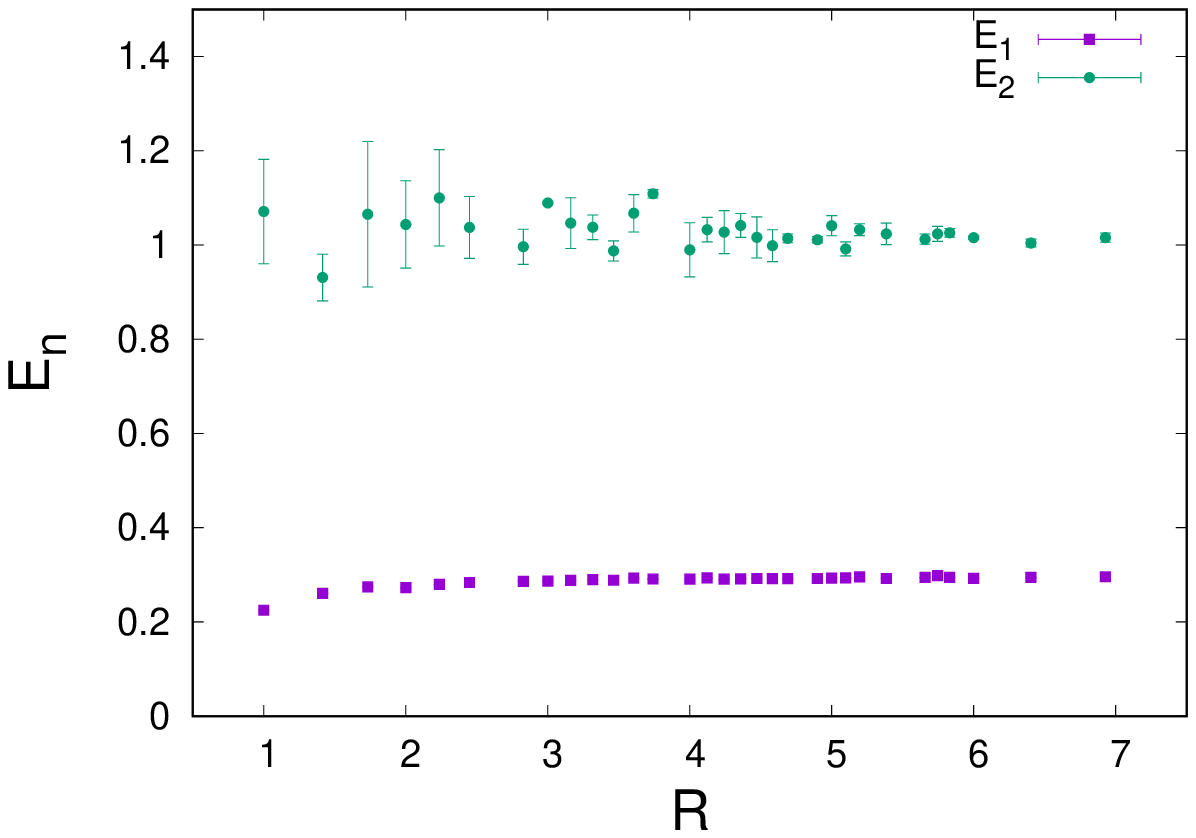}  
}
\caption{(a) An example of fits at $R=6.93$ on a $16^4$ lattice with couplings $\beta=3, \gamma=0.5$. (b) A plot of energy expectation values $E_n(R)$ vs.\ $R$ for $n=1$ and $n=2$. }
\end{figure}
Fig. \ref{fig3} shows an example of these fits at $R=6.93$ on a $16^4$ lattice with couplings $\beta=3, \gamma=0.5$. Fitting through the points at $T=2-5$, we find $E_1=0.2929(6)$ and $E_2(R) = 1.01(1)$. Fig. \ref{fig4} shows energy expectation values $E_n(R)$ vs.\ $R$ for $n=1$ and $n=2$, obtained from a fit to a single exponential. The data and errors were obtained from ten independent runs, each of 77,000 sweeps after thermalization, with data taken every 100 sweeps, computing $\mathcal{T}_{nn}$ from each independent run.

To check a finite size effect we can make the same computation, with the same number of sweeps, only on a $12^4$ lattice.
\begin{figure}[htb]
\center
\subfigure[~]
{
\label{fig5}
\includegraphics[scale=0.5]{fit12.eps}   
}
\subfigure[~]
{
\label{fig6}
\includegraphics[scale=0.5]{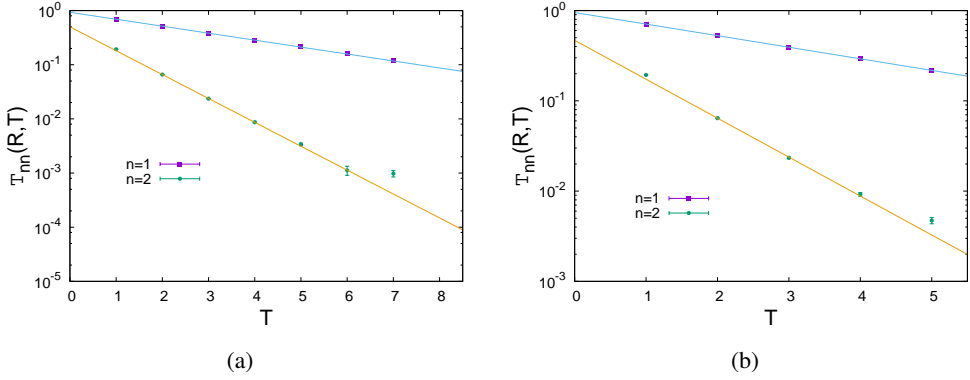}   
}
\caption{(a) An example of fits at $R=6.93$ on a $16^4$ lattice with couplings $\beta=3, \gamma=0.5$ (b) An example of fits at $R=6.93$ on a $12^4$ lattice with couplings $\beta=3, \gamma=0.5$. Showing the fitting examples on two different size lattice next to each other for comparison purpose.}
\end{figure}
Fig. \ref{fig5} and Fig. \ref{fig6} show an example of these fits at $R=6.93$ on a $16^4$ and on a $12^4$ lattice with couplings $\beta=3, \gamma=0.5$ next to each other. Fitting results at $R=6.93$ on a  $12^4$ through the points $T=2-4$ yields $E_2(R)=0.99(2)$. On both lattice volumes the last data point lies a little above the straight line fit, and this is probably a finite size effect.

We also look for any indication of a second stable excited state
\begin{figure}[htb]
\center
\includegraphics[scale=0.5]{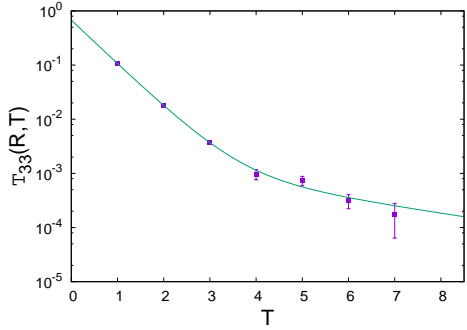}   
\caption{Fitting result for the second stable excited state at fixed $R=6.93$.}
\label{fig7}
\end{figure}
$\mathcal{T}_{33}(R,T)$ vs.\ $T$.  The fit shown is to the sum of exponentials
\beq
          \mathcal{T}_{33}(R,T)  \approx a_1(R) e^{-E_1 T} + a_2(R) e^{-E_2 T} + a_3(R)  e^{-E_3 T} \ ,
\eeq
where $E_1=0.29, E_2=1.02$ are taken from the previous fits.  A sample fit, again at $R=6.93$, is shown in Fig. \ref{fig7}. Obviously one cannot be very impressed by a four parameter fit through a handful of data points, but we show the fit for whatever for it may be worth.

\begin{figure}[htb]
\center
 \includegraphics[scale=0.6]{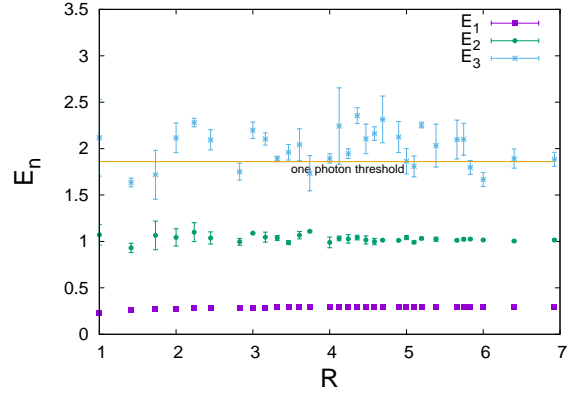}
 \caption{Excitation spectrum in the $q=2$ gauge Higgs theory.}
  \label{fig8}
\end{figure}

In Fig. \ref{fig8}, we show our final result of the excitations spectrums in the $q=2$ gauge Higgs theory.
The one photon threshold is simply $E_1 + m_{\text{photon}} = 0.29 + 1.57(1) = 1.86(1)$ in lattice units.  The important observation is
that $E_2(R)$ lies well below this threshold, which implies that the first excited state of the static fermion-antifermion pair is \emph{stable}.  The second point to note is that $E_3(R)$ seems to lie above or near the one photon threshold.  The indications are that there is no second stable excited state.  States above the first excited state most likely lie above the threshold, and are probably combinations of the ground state plus a massive photon.

\section{Conclusions}

We have presented lattice Monte Carlo evidence for  the existence of a stable excitation of the quantized fields surrounding isolated static charges, in the Higgs phase of the $q=2$ abelian Higgs model in $D=4$ spacetime dimensions. Some obvious questions we may ask next is first, if excitations of this kind are seen in the Abelian Higgs model, would they also be found in non-relativistic models of that kind, so we are considering the application of this kind of analysis to realistic Landau Ginzburg model of superconductivity. Secondly, if there are such excitations, we may ask how they might be observed experimentally in a real superconductor, in ARPES, for example. Finally, we may ask that whether heavy fermions (or even light fermions) have a spectrum of excitations in the electroweak sector of the Standard Model.

\end{document}